\documentclass[%
preprint,
amsmath,assume,
aps,
prb,
]{revtex4-2}

\usepackage{graphicx}
\usepackage{dcolumn}
\usepackage{bm}
\usepackage{here}
\usepackage{amsmath}
\usepackage{color}
\usepackage{ulem}

\newcommand{\keep}[1]{\textcolor{black}{#1}}
\newcommand{\add}[1]{\textcolor{black}{#1}}	
\newcommand{\erase}[1]{\if0{#1}\fi}	

\usepackage{hyperref}

\begin{document}

\preprint{APS/123-QED}

\title{\keep{Fabrication of Oxide/FeSe multilayer films using the PLD technique}}

\author{Tomoki Kobayashi, Hiroki Nakagawa, Ryo Ogawa, and Atsutaka Maeda}
\affiliation{%
 Department of Basic Science, The University of Tokyo, 3-8-1 Komaba, Meguro-ku, Tokyo 153-8902, Japan
}%


\date{\today}

\begin{abstract}

In this study, we demonstrate the successful fabrication of TiO$_2$/FeSe/SrTiO$_3$ and TiO$_2$/FeSe/LaAlO$_3$ heterostructures using PLD. When the growth rates were sufficiently low, anatase TiO$_2$, which has been reported to induce superconductivity in FeSe, could be grown epitaxially on FeSe. \erase{For TiO$_2$/FeSe/SrTiO$_3$, the FeSe/TiO$_2$ interface was relatively clean, as confirmed by the observation}
\erase{ of Laue fringes in FeSe(001) reflection.}
\add{For TiO$_2$/FeSe/SrTiO$_3$ with FeSe layers thicker than 5 nm, the FeSe/TiO$_2$ interface was clean, as confirmed by the observation of Laue fringes around the FeSe(001) reflection.} Although a thinner FeSe layer is expected to enhance $T_{\mathrm{c}}$, 
\erase{our results instead suggest that deposition of TiO$_2$ introduces additional disorder in FeSe. }
\add{we could not observe such behavior because it was difficult to realize interfacial superconductivity while growing anatase TiO$_2$.}
Most strikingly, superconductivity appeared when TiO$_2$ was deposited on FeSe/LaAlO$_3$, clearly demonstrating that the interaction between the top oxide layer and underlying FeSe can induce superconductivity. This work demonstrates the feasibility of fabricating FeSe/oxide superlattices by PLD, establishing a novel platform for the exploration of interfacial superconductivity in iron-based superconductors. 
\end{abstract}

\maketitle



\section{Introduction}
Fe-chalcogenide superconductors have attracted much attention because of their high tunability of the superconducting transition temperature ($T_{\mathrm{c}}$). Although the $T_{\mathrm{c}}$ of bulk FeSe is 9 K at ambient pressure \cite{Hsu_2008}, it can be significantly enhanced by compressive strain \cite{Nabeshima_2018_strain,Nakajima_2021,Ghini_2021}, chemical pressure \cite{Imai_2010}, hydrostatic pressure \cite{Mizuguchi2008,Sun2016}, and electron doping through intercalation \cite{Guo_2010,Shi_2018} or the electric-field effect \cite{Lei_2016,Shiogai_2016,Wang_2016,Hanzawa_2016,Kouno_2018,Shikama_2020,Shikama_2021}. In particular, a zero-resistance temperature ($T_{\mathrm{c}}^{\mathrm{zero}}$) as high as 46 K has been achieved in Refs. \cite{Shikama_2020,Shikama_2021}.

Intriguingly, ultrathin FeSe films grown on typical oxide substrates such as SrTiO$_3$ (STO) \cite{Wang_2012} and TiO$_2$ \cite{Ding_2016,Rebec_2017} also exhibit enhanced superconductivity owing to interfacial effects. Electron transfer \cite{Liu_2012,Zhang_2017,Zhao_2018} and interfacial electron–phonon coupling \cite{Lee_2014,Zhang_2016_Kdose,Song_2019} have been reported to play essential roles in this enhancement. For FeSe/STO, angle-resolved photoemission spectroscopy (ARPES) measurements revealed a gap opening at the Fermi level as high as 65 K \cite{He_2013}. However, it remains unclear whether this observation is a direct signature of superconductivity, since transport measurements indicate an onset $T_{\mathrm{c}}$ ($T_{\mathrm{c}}^{\mathrm{onset}}$) around 40 K, which is much lower than the gap-opening temperature \cite{Wang_2012, Pedersen_2020, Faeth_2021}. Furthermore, the resistance drop is typically broad, and the zero-resistance state emerges only at $T_{\mathrm{c}}^{\mathrm{zero}}=$ 10–30 K, well below $T_{\mathrm{c}}^{\mathrm{onset}}$. This unexpectedly low $T_{\mathrm{c}}^{\mathrm{zero}}$ has been attributed to the Berezinskii–Kosterlitz–Thouless (BKT) transition\cite{Wang_2012} or strong inhomogeneity effects\cite{Jiao_2023}, both of which originate from the two-dimensional (2D) nature of superconductivity in FeSe/STO. Therefore, suppressing the 2D character while keeping the presence of the interfaces is a promising approach to achieving a higher $T_{\mathrm{c}}^{\mathrm{zero}}$ in this system.
One possible strategy is to fabricate a superlattice composed of alternating FeSe and oxide layers. In such a structure, if the intermediate FeSe layers are thinner than the coherence length, superconductivity at each FeSe/oxide interface may \erase{couple}\add{be Josephson-coupled} and reduce the 2D nature. In addition, interfacial electron–phonon coupling may be enhanced, leading to a higher $T_{\mathrm{c}}$ \cite{Coh_2016}. So far, superlattices based on FeSe have been fabricated using FeSe and FeTe\cite{Nabeshima_2017_superlattice}, which are chemically and structurally similar compounds. However, superlattices composed of FeSe and oxide materials are more challenging, because direct growth of oxide layers on FeSe is difficult due to its high sensitivity to oxygen. As a first step toward realizing such superlattices, it is important to construct oxide/FeSe/oxide multilayer structures(Fig. \ref{Fig:Fig1}(a)), in which superconducting layers are confined near each interface. 
\add{Even in this structure, superconductivity originating at the two interfaces can be coupled to form a Josephson junction, leading to the suppression of 2D nature and an enhancement of $T_{\mathrm{c}}$ if the FeSe interlayer is sufficiently thin.}

Pulsed laser deposition (PLD) offers a promising route to deposit oxide materials without oxidizing FeSe. In the PLD growth process, a high-energy laser pulse ablates the target material and generates vaporized species that retain the stoichiometry of the target. As a result, PLD can produce thin films that nearly reproduce the target composition\cite{Lowndes_1996}. This feature of PLD allows the growth of epitaxial oxide films under vacuum conditions. Indeed, growth of STO under vacuum has been achieved by tuning laser condition \cite{Lee_2016}. Furthermore, different materials can be deposited simply by switching the target, which is advantageous for exploring high-$T_{\mathrm{c}}$ superconductivity in FeSe/oxide systems via interface engineering.

In this study, we demonstrate the successful fabrication of TiO$_2$/FeSe/STO and TiO$_2$/FeSe/LaAlO$_3$ heterostructures using PLD. When the growth rates were sufficiently low, anatase TiO$_2$, which has been reported to induce superconductivity in FeSe \cite{Ding_2016}, could be grown epitaxially on FeSe. For TiO$_2$/FeSe/STO, the FeSe/TiO$_2$ interface was relatively clean, as confirmed by the observation of Laue fringes in FeSe(001) reflection. With FeSe thicknesses $d_{\mathrm{FeSe}} > 5$ nm, no further enhancement of superconductivity was observed, probably because the observed $T_{\mathrm{c}}$ originates from the FeSe/STO interface. Although a thinner FeSe layer is expected to enhance $T_{\mathrm{c}}$, \erase{we observed suppression of superconductivity due to deterioration during TiO$_2$ deposition. }
\add{we could not observe such behavior because it was difficult to realize interfacial superconductivity while growing anatase TiO$_2$.}
Most strikingly, however, superconductivity appeared when TiO$_2$ was deposited on FeSe/LaAlO$_3$, clearly demonstrating that the interface between the top oxide layer and underlying FeSe can induce superconductivity. This work demonstrates the feasibility of fabricating FeSe/oxide superlattice by PLD, establishing a novel platform for the exploration of interfacial superconductivity in iron-based superconductors.


\section{EXPERIMENTAL METHODS}
We used STO and LaAlO$_3$(LAO) substrates for the growth of oxide/FeSe/oxide heterostructures. 
Prior to PLD deposition, surface treatments were performed on both substrates. 
STO substrates were annealed in air and subsequently soaked in water, following the detailed procedure described in previous reports \cite{Kobayashi_2022}. 
For LAO, the substrates were first annealed at 1000 °C for 12 hours in air, followed by quenching from 500 °C to room temperature ($\sim 27^\circ$C) during cooling to obtain an atomically flat surface. 
After quenching, the substrates were soaked in deionized water for two days to achieve AlO$_2^{-}$ termination \cite{Kim_2019}. 
FeSe films were grown on STO and LaAlO$_3$ substrates at a substrate temperature of 500 °C with a laser repetition rate of 1 Hz under high vacuum ($\sim 1.0 \times 10^{-8}$ Torr). 
As the top oxide material, we employed TiO$_2$. 
The anatase phase of TiO$_2$ has an in-plane lattice constant comparable to that of FeSe ($a = 3.78$ Å for TiO$_2$ and 3.81 Å for FeSe/STO), and it has been reported that FeSe on anatase TiO$_2$ exhibits superconductivity similar to that observed for FeSe/STO \cite{Ding_2016}. 
After the growth of FeSe, TiO$_2$ layers were subsequently deposited at the same substrate temperature under high vacuum ($\sim 1.0 \times 10^{-8}$ Torr), using a pellet fabricated from rutile TiO$_2$ powder. 
All films were finally capped with \add{undoped} Si to prevent degradation due to air exposure.
\add{We note that this Si capping layer does not change the carrier density of TiO$_2$ or FeSe because it is an intrinsic semiconductor with a very low carrier density.}
The orientation and crystal structure of the grown films were characterized by X-ray diffraction (XRD) using Cu K$\alpha$ radiation at room temperature. 
\erase{Film thicknesses were estimated either from Laue fringes around the (001) reflection peak or}
\erase{with a Dektak 6M stylus profiler. }
\add{Film thicknesses were estimated from the Laue fringes around the FeSe(001) reflections or from the deposition time. Details are described in the Supplemental Material\cite{SM}.}
Resistivity measurements were carried out using a Physical Property Measurement System (PPMS) in the temperature range of 2–300 K.


\section{Results}

\begin{figure*}[htb]
\centering
\includegraphics[width=\linewidth]{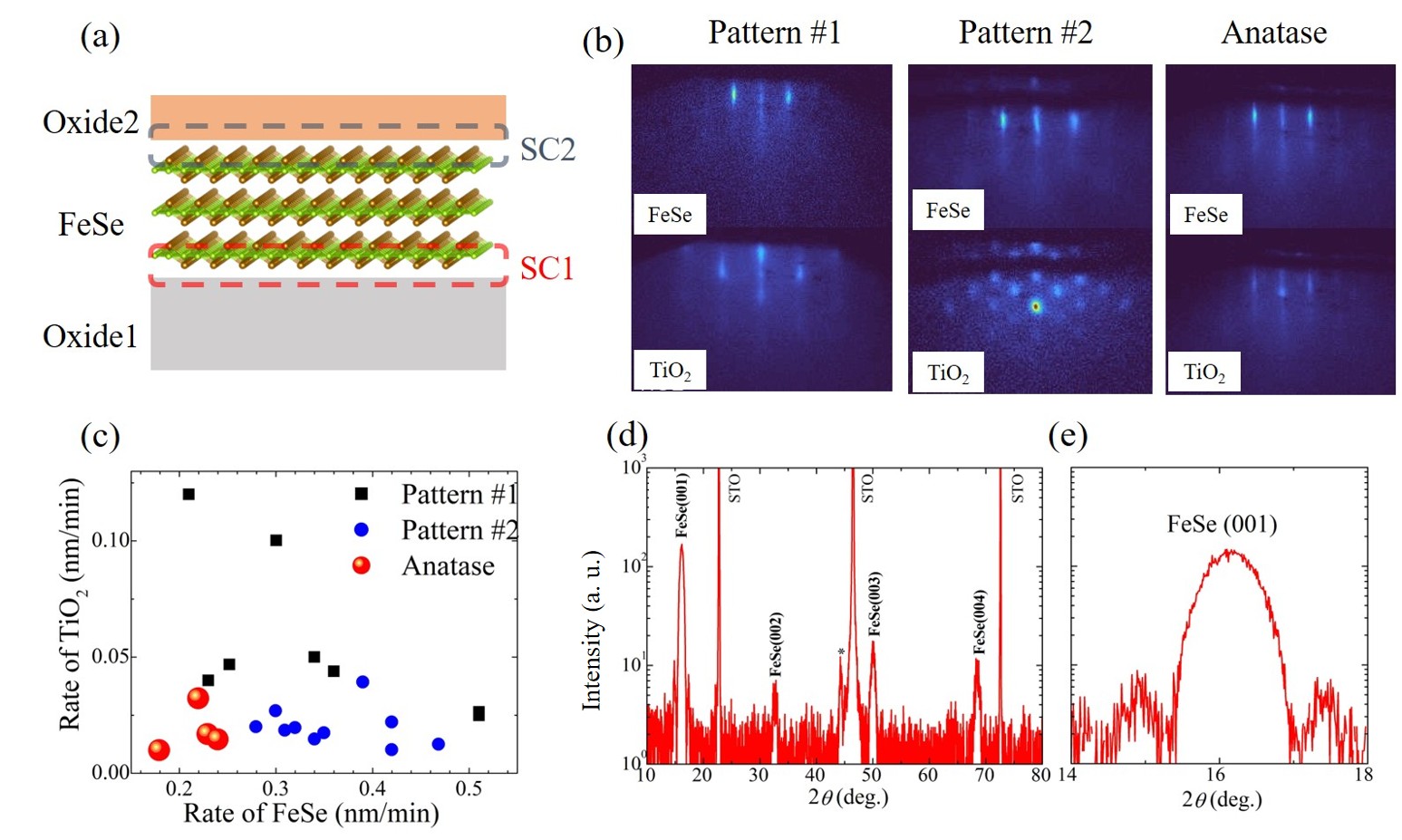}
\caption{\label{Fig:Fig1} 
(a) Schematic illustration of the oxide/FeSe/oxide heterostructure. 
Superconductivity originating from each FeSe/oxide interface emerges. 
(b) RHEED images obtained after the growth of FeSe (top) and TiO$_2$ (bottom). 
Left, middle, and right images correspond to $r_{\mathrm{TiO_2}} > 0.04$ nm/min, 
$r_{\mathrm{TiO_2}} < 0.04$ nm/min with $r_{\mathrm{FeSe}} > 0.25$ nm/min, 
and $r_{\mathrm{TiO_2}} < 0.04$ nm/min with $r_{\mathrm{FeSe}} < 0.25$ nm/min, respectively. 
(c) Summary of RHEED patterns for various $r_{\mathrm{TiO_2}}$ and $r_{\mathrm{FeSe}}$. 
(d) XRD pattern of a representative TiO$_2$/FeSe/STO heterostructure 
with $d_{\mathrm{FeSe}} = $\ \erase{5}\add{10} nm and $d_{\mathrm{TiO_2}} \sim 2$ nm. 
\add{The peak indicated by an asterisk is the Fe(101) peak originating from the sample stage.}
(e) Enlarged plot of the FeSe(001) reflection.
}
\end{figure*}

The growth of anatase TiO$_2$ on FeSe requires careful optimization of deposition parameters. 
Previous studies of PLD growth of anatase TiO$_2$ on LAO showed that the deposition rate influences the density of oxygen vacancies \cite{Tachikawa_2012}, which may affect the growth of anatase phase on FeSe. 
We fabricated TiO$_2$/FeSe/STO heterostructures while varying the deposition rates of both FeSe ($r_{\mathrm{FeSe}}$) and TiO$_2$ ($r_{\mathrm{TiO_2}}$). 
The phase of the top TiO$_2$ layer was examined using Reflection High-Energy Electron Diffraction (RHEED) just after the deposition (Fig. \ref{Fig:Fig1}(b)). 
Because the in-plane lattice constants of FeSe and anatase TiO$_2$ are nearly identical, the streak spacing in the RHEED patterns is expected to be equivalent after FeSe deposition and after deposition of the top TiO$_2$ layer. 
When TiO$_2$ was deposited at a rate exceeding 0.04 nm/min, the streak spacing clearly deviated from that of FeSe, irrespective of the FeSe growth rate, indicating that the top oxide layer was not anatase TiO$_2$. 
Furthermore, we found that when FeSe itself was grown at a rate higher than 0.25 nm/min, the subsequent TiO$_2$ layer failed to crystallize in the anatase phase, even when the TiO$_2$ deposition rate was kept below 0.04 nm/min. 
These observations suggest that both the growth rates of FeSe and TiO$_2$ must be carefully controlled to suppress competing phases.

Successful growth of anatase TiO$_2$ was achieved only under the combined conditions of $r_{\mathrm{FeSe}} < 0.25$ nm/min and $r_{\mathrm{TiO_2}} < 0.04$ nm/min, as illustrated in the right panel in Fig. \ref{Fig:Fig1}(b). 
The corresponding RHEED image exhibits streaks with identical spacing for FeSe and TiO$_2$, confirming that the grown oxide phase was anatase. 
\add{The appropriateness of this low $r_{\mathrm{TiO_2}}$ was confirmed by growing a TiO$_2$ film on an LAO substrate under similar conditions, where the anatase phase was confirmed by RHEED and XRD measurements (see the Supplemental Material\cite{SM}).}
A systematic summary of the RHEED images obtained at different growth rates is also provided in Fig. \ref{Fig:Fig1}(c), clearly indicating that both $r_{\mathrm{TiO_2}}$ and $r_{\mathrm{FeSe}}$ are important to obtain the anatase phase on FeSe.
Figure \ref{Fig:Fig1}(d) shows the XRD pattern of a representative anatase TiO$_2$/FeSe/STO heterostructure.
Distinct FeSe(00$l$) diffraction peaks were observed even after deposition of the TiO$_2$, demonstrating that the crystalline order of FeSe was preserved during the process. 
In addition, Laue fringes were clearly visible around the FeSe(001) reflection (Fig. \ref{Fig:Fig1}(e)), indicating flat interfaces between FeSe and TiO$_2$. 
These results demonstrate that careful control of the growth rates ($r_{\mathrm{TiO_2}}$ and $r_{\mathrm{FeSe}}$) enables the fabrication of TiO$_2$/FeSe/STO heterostructures, while preserving the single crystallinity of FeSe and ensuring flat top oxide/superconductor interfaces.

\begin{figure}[htpb]
\centering
\includegraphics[width=\linewidth]{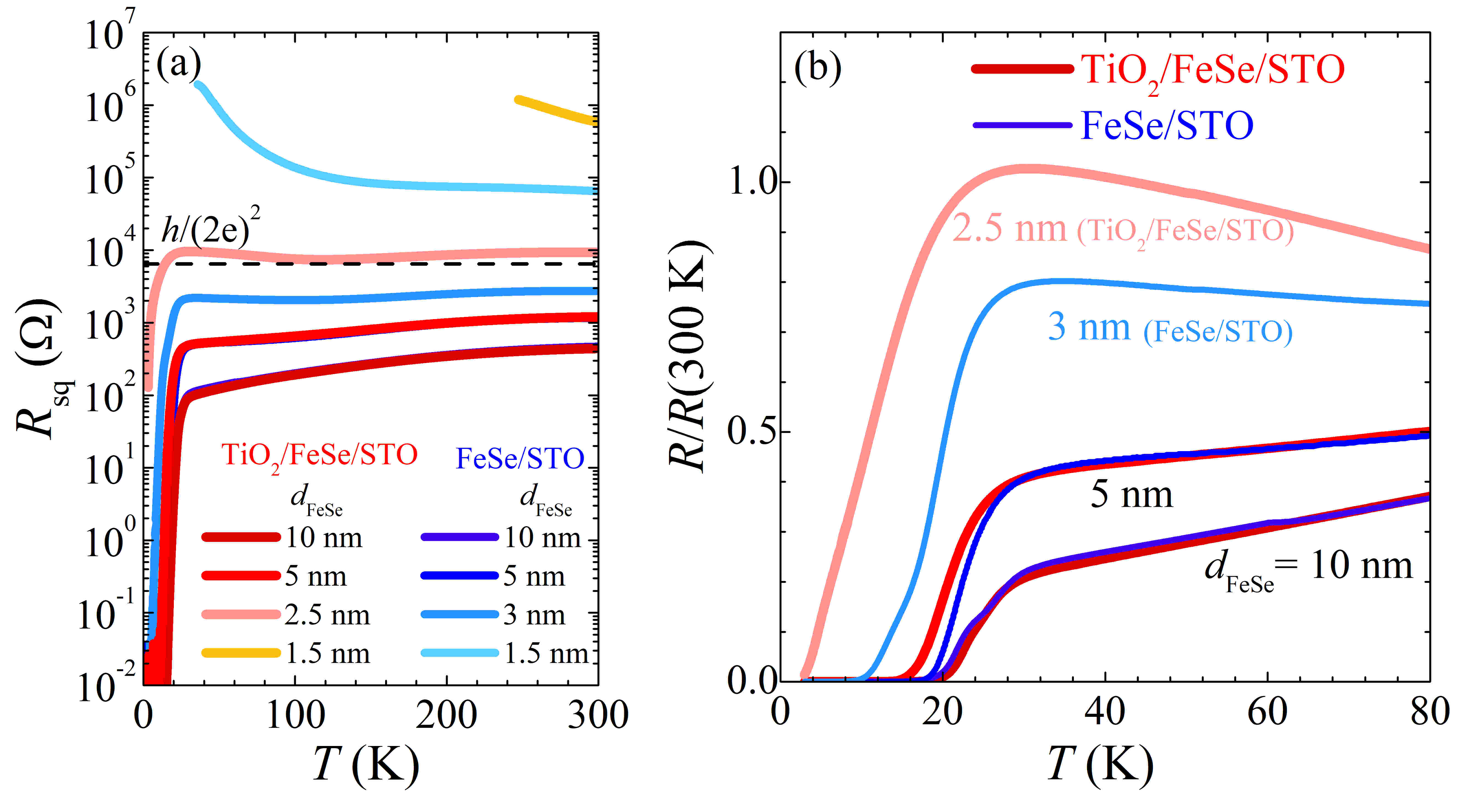}
\caption{\label{Fig:Fig2} 
(a) Temperature dependence of the sheet resistance of TiO$_2$/FeSe/STO and FeSe/STO films with different $d_{\mathrm{FeSe}}$. 
For TiO$_2$/FeSe/STO heterostructures, the TiO$_2$ thickness was kept at $d_{\mathrm{TiO_2}} \sim 2$ nm. 
(b) Temperature dependence of the normalized resistance of TiO$_2$/FeSe/STO and FeSe/STO films in the range 0–80 K.
}
\end{figure}

Figure \ref{Fig:Fig2}(a) shows the temperature dependence of the sheet resistance of TiO$_2$/FeSe/STO heterostructures and FeSe/STO films with FeSe thicknesses ($d_{\mathrm{FeSe}}$) ranging from 10 to \erase{2}\add{1.5} nm, while keeping the TiO$_2$ thickness ($d_{\mathrm{TiO_2}}$) fixed at 2 nm. 
When $d_{\mathrm{FeSe}} = 10$ \add{ and 5 } nm, \add{the sheet resistance and residual resistance ratio $\mathrm{RRR} (\ =R_{(300 \mathrm{K})}/R_{(40 \mathrm{K})}) $ were almost the same for the TiO$_2$/FeSe and FeSe/STO films. This suggests that TiO$_2$ deposition did not cause significant degradation of FeSe, which is consistent with the presence of Laue fringes around the FeSe(001) reflection.} \add{Furthermore,} $T_{\mathrm{c}}^{\mathrm{onset}}$ was unchanged regardless of the presence of the TiO$_2$ overlayer (Fig. \ref{Fig:Fig2}(b)). 
This result is not surprising, since possible superconductivity at the TiO$_2$/FeSe and FeSe/STO interfaces is not expected to couple owing to the short coherence length \cite{Kobayashi_2023}. 
This indicates that the observed superconductivity originates from the FeSe/STO interface. 
For thinner FeSe layers, \erase{($d_{\mathrm{FeSe}} < 3$ nm), the behavior differed between films with and without
the TiO$_2$ overlayer.} 
\add{superconductivity in FeSe/STO was slightly suppressed as the FeSe thickness decreased, and the 1.5-nm-thick film did not exhibit superconductivity. This suppression of superconductivity is in contrast to the literature\cite{Wang_2015,Zhao_2018} and our previous study\cite{Kobayashi_2022,Kobayashi_2023}, which reported an enhancement of superconductivity with decreasing thickness. This discrepancy may be due to the FeSe deposition rate being as low as 0.1 nm/min in the present study, whereas it was 0.2--0.5 nm/min in the previous study\cite{Kobayashi_2022}. Although such a low deposition rate was necessary for the subsequent growth of anatase TiO$_2$, it was not optimal for obtaining high-quality ultrathin FeSe films.}
\add{When comparing the TiO$_2$/FeSe ($d_{\mathrm{FeSe}}=2.5$ nm) and FeSe ($d_{\mathrm{FeSe}}=3$ nm) films, the sheet resistance increased by a factor of 3.4 and the transition became broader after TiO$_2$ deposition.}
At $d_{\mathrm{FeSe}} \sim 1.5$ nm, 
\erase{a similar increase in sheet resistance was observed, but }
\add{TiO$_2$/FeSe/STO exhibited higher resistance than FeSe/STO.}
\erase{the}\add{The} resistance \add{of these films }exceeded the quantum critical value ($h/(2e)^2$) associated with the superconductor–insulator transition\cite{Goldman_1998}, and no superconducting transition was detected. 
\add{These differences between films with and without a TiO$_2$ layer may be caused either by a slight difference in thickness or by the deposition of TiO$_2$ on FeSe.
In the latter case, one possible mechanism is structural damage to, or oxidation of, FeSe caused by high-energy species in the laser ablation plume.}
\erase{Although an enhancement of $T_{\mathrm{c}}$ might be expected for such thin FeSe films, these}
\erase{results instead suggest that deposition of TiO$_2$ introduces additional disorder in FeSe.}

\begin{figure}[htpb]
\centering
\includegraphics[width=\linewidth]{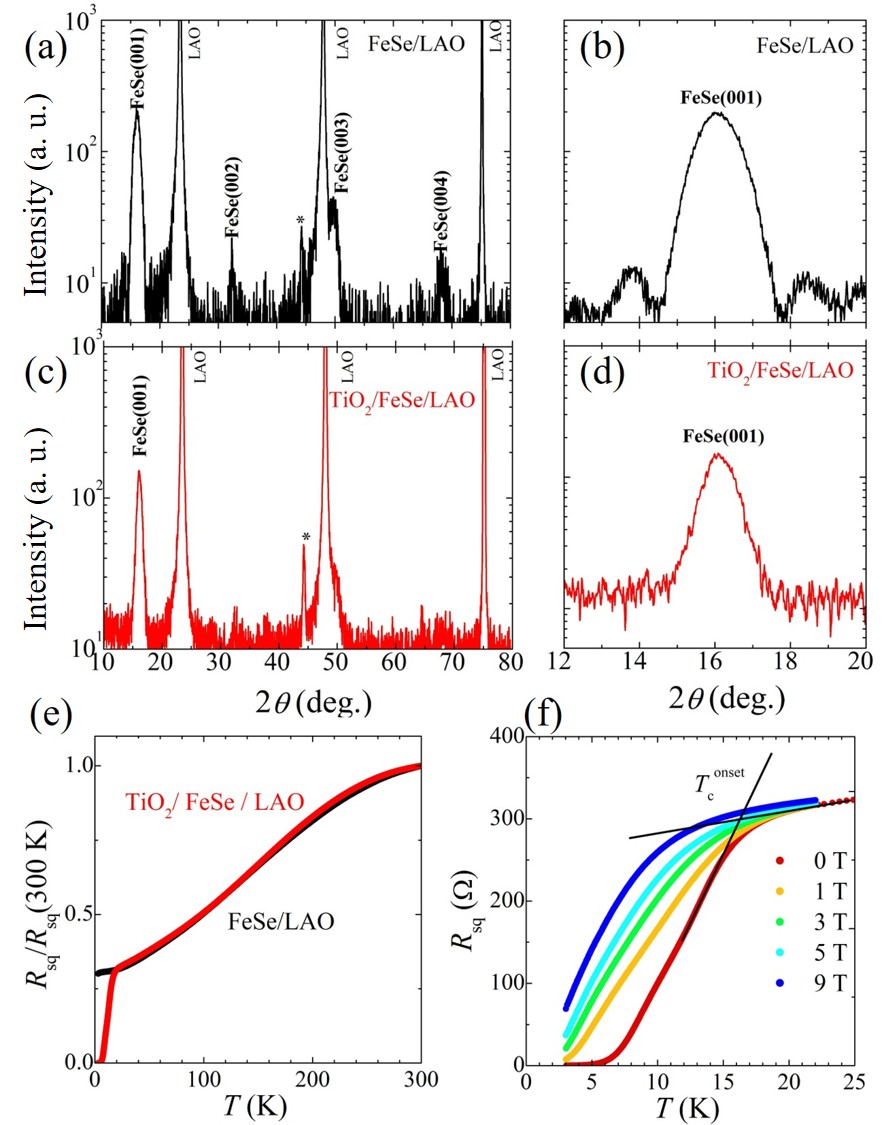}
\caption{\label{Fig:Fig3} 
(a) and (c) XRD patterns of FeSe/LAO and TiO$_2$/FeSe/LAO, respectively. (b) and (d) show enlarged plots around the FeSe(001) reflections.
(e) Temperature dependence of the normalized resistance of TiO$_2$/FeSe/LAO and FeSe/LAO films with $d_{\mathrm{FeSe}} \sim 5$ nm. 
(f) Temperature dependence of the sheet resistance of TiO$_2$/FeSe/LAO films under magnetic fields of 0--9 T.
}
\end{figure}

To overcome this issue, we employed a different substrate material, LAO, in an attempt to prepare interfaces with reduced degradation. 
LAO has a lattice constant of $a = 3.78$ \AA, which is almost identical to that of anatase TiO$_2$.
Therefore, using an LAO substrate is expected to provide a better interface, as the in-plane lattice constants of anatase TiO$_2$ and FeSe are more closely matched. 
Figures \ref{Fig:Fig3}(a) and (c) show the XRD patterns of FeSe/LAO ($d_{\mathrm{FeSe}}=5.5$ nm)and TiO$_2$/FeSe/LAO ($d_{\mathrm{FeSe}}=5.5$ nm, $d_{\mathrm{TiO_2}} \sim 2$ nm) . 

\erase{All FeSe (00$l$) peaks were observed for FeSe/LAO; however, only the FeSe(001) reflection} 
\erase{ was observed for TiO$_2$/FeSe/LAO. }
\add{All FeSe (00$l$) peaks were observed for FeSe/LAO; however, the FeSe(002) and (004) reflections were not observed for TiO$_2$/FeSe/LAO because of their reduced intensities. }
Furthermore, Laue fringes were not observed for the film with a TiO$_2$ overlayer (Fig. \ref{Fig:Fig3}(d)), in contrast to FeSe/LAO (Fig. \ref{Fig:Fig3}(b)).
These results indicate that the TiO$_2$/FeSe interface became unexpectedly rough and that deposition of TiO$_2$ probably introduces disorder into FeSe, as in the case of FeSe/STO.
However, in contrast to the films on STO, emergence of superconductivity only after depositing TiO$_2$ was observed for the films on LAO.
Figure \ref{Fig:Fig3}(e) shows the temperature dependence of the resistance of FeSe/LAO films with and without a TiO$_2$ overlayer.
FeSe/LAO without TiO$_2$ did not show a superconducting transition down to 2 K.
\add{This absence of superconductivity, unlike in FeSe/STO, can be attributed to the lack of electron doping from the substrate. First, the work function of AlO$^{_2^-}$-terminated LAO is calculated to be 6.87 eV\cite{Ryan_2016}, which is higher than that of FeSe (5.1 eV), in contrast to the situation in FeSe/STO\cite{Zhang_2017}.}
\add{Second, although charge transfer driven by polar discontinuity, as in LAO/STO, may in principle occur in LAO-based structures, the B-terminated surface employed in this study would result in hole doping rather than electron doping.}
In contrast to FeSe/LAO, the film with a TiO$_2$ overlayer exhibited a superconducting transition at low temperatures. 
The superconducting origin of the resistive drop was further confirmed by its suppression under applied magnetic fields (Fig. \ref{Fig:Fig3}(f)). 
The onset transition temperature was $T_{\mathrm{c}}^{\mathrm{onset}} = 16$ K, which is higher than that of bulk FeSe. 
\add{Zero resistivity was observed at 3 K, which may be related to structural damage, as suggested by the absence of Laue fringes around the FeSe(001) reflection.}
This result suggests that the emergence of superconductivity can be attributed to interfacial effects at the TiO$_2$/FeSe interface. 
\add{The detailed mechanism underlying the emergence of superconductivity is still unclear. However, we believe that electron transfer from the oxide to FeSe is likely to be relevant, similar to the case of FeSe/STO. This possibility could be investigated by electron energy-loss spectroscopy (EELS) combined with scanning transmission electron microscopy (STEM)\cite{Zhao_2018}.} 
The different behavior of the films on STO and LAO upon TiO$_2$ deposition is likely related to the presence of superconductivity originating from the FeSe/substrate interface. 
In the case of FeSe/STO, superconductivity with $T_{\mathrm{c}}^{\mathrm{onset}} \sim 25$ K emerges from the FeSe/STO interface and may mask a possible superconducting transition originating from the TiO$_2$/FeSe interface with a lower $T_{\mathrm{c}}$, whereas FeSe/LAO does not exhibit such substrate-induced superconductivity.

The emergence of superconductivity in TiO$_2$/FeSe/LAO demonstrates that a top oxide overlayer can induce the emergence of superconductivity, analogous to the substrate-induced effect\cite{Ding_2016}. 
This observation indicates that coupling between the superconductivity originating at each interface in the TiO$_2$/FeSe/oxide heterostructure is, in principle, possible. 
Therefore, our results suggest that constructing a TiO$_2$/FeSe superlattice provides a promising platform for enhancing or engineering interfacial superconductivity. 
At this stage, however, \erase{degradation of FeSe during TiO$_2$ deposition makes the coupling of superconductivity difficult.}
\add{the existence of superconductivity originating at the TiO$_2$/FeSe interface on the STO substrate was not experimentally confirmed because the $T_{\mathrm{c}}$ associated with FeSe/STO may be higher than that associated with the TiO$_2$/FeSe interface.}
\add{To reveal the existence of superconductivity near the TiO$_2$/FeSe interface, AC diamagnetic measurements would be promising. However, because the superconductivity is expected to be inhomogeneous, improving the interface quality is also necessary.}
\add{In addition, degradation of FeSe during TiO$_2$ deposition may hinder coupling between the superconducting regions at the two interfaces.} 
\add{In this case, }
\erase{One}\add{one} possible degradation mechanism is damage to FeSe caused by high-energy species in the laser ablation plume.
Thus, a promising way to mitigate damage is to reduce the kinetic energy of the plume.
This could be achieved by introducing a He buffer gas during PLD\cite{Takahashi_2021}, which reduces the kinetic energy through scattering between plume species and He atoms, and will be explored in future work.

\section {CONCLUSION}

In this study, we demonstrate the successful fabrication of TiO$_2$/FeSe/STO and TiO$_2$/FeSe/LAO heterostructures using PLD. When the growth rates were sufficiently low, anatase TiO$_2$, which has been reported to induce superconductivity in FeSe, could be grown epitaxially on FeSe. 
\add{For TiO$_2$/FeSe/SrTiO$_3$ with FeSe layers thicker than 5 nm, the FeSe/TiO$_2$ interface was clean, as confirmed by the observation of Laue fringes around the FeSe(001) reflection.} Although a thinner FeSe layer is expected to enhance $T_{\mathrm{c}}$, 
\erase{our results instead suggest that deposition of TiO$_2$ introduces additional disorder in FeSe. }
\add{we could not observe such behavior because it was difficult to realize interfacial superconductivity while growing anatase TiO$_2$.}
Most strikingly, however, superconductivity appeared when TiO$_2$ was deposited on FeSe/LAO, clearly demonstrating that the interface between the top oxide layer and underlying FeSe can induce superconductivity. This work demonstrates the feasibility of fabricating FeSe/oxide superlattices by PLD, establishing a novel platform for the exploration of interfacial superconductivity in iron-based superconductors. 

\section {ACKNOWLEDGEMENT}
We would like to acknowledge K Ueno and H Okuma at the University of Tokyo for their technical support of the XRD measurements.
This research was supported by JSPS KAKENHI Grants No. JP24KJ0799 and JP24K06952. 
\bibliography{reference}

\end{document}